\documentstyle[aps,eqsecnum,amsfonts,psfig,preprint]{revtex}

\begin{document}

\title{ \hfill OKHEP-95-01\\ \hfill Imperial/TP/94--95/23\\
Finite-Element Lattice Hamiltonian Matrix Elements.
Anharmonic Oscillators}
\author{Kimball A. Milton\thanks{E-mail: kmilton@uoknor.edu or
k.milton@ic.ac.uk (until 1 July 1995)}}
\address{\it Department of Physics and Astronomy,
The University of Oklahoma, Norman OK 73019, USA\thanks{
Permanent address}
\\
Theoretical Physics Group, Blackett Laboratory,
 Imperial College, Prince Consort Road, London
SW7 2BZ, UK}
\author{Rhiju Das\thanks{E-mail: rdas@bashful.ossm.edu}}
\address{Oklahoma School of Science and Mathematics, 1141
Lincoln Blvd.,  Oklahoma City, OK 73104 USA}
\date{\today}
\maketitle

\begin{abstract}
The finite-element approach to lattice field theory is both
highly accurate (relative errors $\sim 1/N^2$, where $N$ is
the number of lattice points) and exactly unitary (in the
sense that canonical commutation relations are exactly preserved
at the lattice sites).  In this paper we construct matrix elements
for the time evolution operator for the anharmonic oscillator,
for which the continuum Hamiltonian is $H=p^2/2+\lambda q^{2k}/2k$.
Construction of such matrix elements does not require solving
the implicit equations of motion.  Low order approximations turn
out to be quite accurate.  For example, the matrix element
of the time evolution operator in the harmonic oscillator
ground state gives a result for the $k=2$ anharmonic oscillator
ground state energy accurate to better than 1\%, while a two-state
approximation reduces the error to less than 0.1\%.  Accurate
wavefunctions are also extracted.  Analogous results may be
obtained in the continuum, but there the computation is more
difficult, and not generalizable to field theories in more
dimensions.
\end{abstract}
\section{Introduction}

For over a decade now, the finite-element method has been developed
for
application to quantum systems.  (For a review of the program see
\cite{review}.)  The essence of the approach is to put the Heisenberg
equations of motion for the quantum system on a Minkowski space-time
lattice
in such a way as to preserve exactly the canonical commutation
relations
at each lattice site. Doing so  corresponds precisely to the
classical
finite-element prescription of requiring continuity at the lattice
sites
while imposing the equations of motion at the Gaussian knots, a
prescription
chosen to minimize numerical error.  We have applied this technique
to
examples in quantum mechanics and to quantum field theories in two
and four
space-time dimensions.  In particular, recent work has concentrated
on
Abelian and non-Abelian gauge theories
\cite{nagt,sing,mmsb,mmsb2,twodqed},
especially on issues of chiral symmetry breaking.

Because it is the equations of motion that are discretized, a
lattice Lagrangian does not exist in Minkowski space.
This is because the equations of motion are in general nonlocal,
involving fields at all
previous (but not later) times.  Similarly, a lattice Hamiltonian
does
not exist, in the sense of an operator from which the equations of
motion can be derived.

However, because the formulation is  unitary, a unitary
time-evolution
operator must exist which carries fields from one lattice time to
the next.
For linear finite elements this operator in quantum mechanics
 has been explicitly constructed
\cite{dtqm}.  Construction of this operator requires solving the
equations
of motion, which are implicit.  Therefore, it is most useful, and
perhaps
surprising, that when  matrix elements of the time
evolution operator are constructed in a harmonic oscillator basis,
they do not require the solution
of the equations of motion \cite{bss}.  Although these general
formulas were
derived some years ago, it seems they have not been exploited.
Our purpose here is to study, in a simple context, the matrix
elements of the evolution operator, and see how accurately
spectral information and wavefunctions may be extracted.\footnote{
A preliminary version of this work appears in \cite{coc}.}
Our goal, of course, is to apply similar techniques
in gauge theories, for example, to study chiral symmetry breaking
in QCD.

\section{Review of the Finite-Element Method}

Let us consider a quantum mechanical system with one degree of
freedom
governed by the continuum Hamiltonian
\begin{equation}
H={p^2\over2}+V(q),
\label{ham}
\end{equation}
from which follow the Heisenberg equations
\begin{equation}
\dot p=-V'(q),\quad \dot q=p.
\label{eqm}
\end{equation}
These equations are to be solved subject to the initial condition
\begin{equation}
[q(0),p(0)]=i.
\end{equation}
It immediately follows from (\ref{eqm}) that the same relation
holds at any later time
\begin{equation}
[q(t),p(t)]=i.
\end{equation}

Now suppose we introduce a time lattice by subdividing the interval
$(0,T)$ into $N$ subintervals each of length $h$.  On each
subinterval
(``finite element'') we express the dynamical variables as $r$th
degree
polynomials
\begin{equation}
p(t)=\sum_{k=0}^r a_k(t/h)^k,\quad q(t)=\sum_{k=0}^r b_k(t/h)^k,
\end{equation}
where $t$ is a local variable ranging from $0$ to $h$.  We
determine the
$2(r+1)$ operator coefficients $a_k$, $b_k$, as follows:
\begin{enumerate}
\item On the first finite element let
\begin{equation}
a_0=p_0=p(0),\quad b_0=q_0=q(0).
\end{equation}
\item Impose the equations of motion (\ref{eqm}) at $r$ points
within the
finite element, at $\alpha_i h$, $i=1,2,\dots,r$, where
$0<\alpha_1<\alpha_2<
\cdots<\alpha_r<1$.
  This then gives
\begin{equation}
p(h)\approx p_1=\sum_{k=0}^r a_k,\quad q(k)\approx q_1=
\sum_{k=0}^r b_k.
\end{equation}
\item Proceed to the next finite element by requiring continuity
(but not
continuity of derivatives) at the lattice sites, that is, on the
second
finite element, set
\begin{equation}
a_0=p_1,\quad b_0=q_1,
\end{equation}
and again impose the equations of motion at $\alpha_ih$, and so on.
\end{enumerate}
How are the $\alpha_i$'s determined?  By requiring preservation of
the
canonical commutation relations at each lattice site,
\begin{equation}
[q_1,p_1]=[q_0,p_0]=i,
\label{unitarity}
\end{equation}
one finds
\begin{mathletters}
\begin{eqnarray}
r=1 \quad (\mbox{linear finite elements}):
\qquad \alpha&=&{1\over2}\\
r=2 \quad (\mbox{quadratic finite elements}): \qquad
\alpha_\pm&=&{1\over2}\pm
{1\over2\sqrt{3}}\\
r=3 \quad (\mbox{cubic finite elements}): \qquad
\alpha_{1,3}&=&{1\over2}\mp{\sqrt{3}\over
2\sqrt{5}},\quad \alpha_2={1\over2}
\end{eqnarray}
\end{mathletters}
These points are exactly the Gaussian knots, that is,
the roots of the
$r$th Legendre polynomial,
\begin{equation}
P_r(2\alpha-1)=0.
\end{equation}
Amazingly, these are precisely the points at which the numerical
error
is minimized.  It is known for classical equations that if one
uses
$N$ $r$th degree finite elements the relative error goes like
$N^{-2r}$,
while imposing the equations at any other points would give errors
like $N^{-r}$.

Let us consider a simple example.  The quartic anharmonic oscillator
has the continuum Hamiltonian
\begin{equation}
H={1\over2} p^2+{1\over4}\lambda q^4,
\label{qaho}
\end{equation}
for which the equations of motion are
\begin{equation}
\dot q=p,\quad \dot p=-\lambda q^3.
\end{equation}
If we use the linear ($r=1$) finite-element prescription given
above,
the corresponding discrete lattice equations are
\begin{equation}
{q_1-q_0\over h}={p_1+p_0\over2},\quad {p_1-p_0\over
h}=-{\lambda\over8}
(q_1+q_0)^3.
\label{leqm}
\end{equation}
(Notice the easily remembered mnemonic for linear finite elements:
Derivatives are replaced by forward differences, while
undifferentiated
operators are replaced by forward averages.)
By commuting the first of these equations with $p_1+p_0$ and
the second
with $q_1+q_0$ the unitarity condition (\ref{unitarity}) follows
immediately.  These equations are implicit, in the sense that we
must solve a nonlinear equation to find $q_1$ and $p_1$ in terms
of $q_0$ and $p_0$.  Although such a solution can be given, let
us make a simple approximation, by expanding the dynamical operators
at time 1 in powers of $h$, with operator coefficients at time 0.
Those coefficients are determined by (\ref{leqm}), and a very simple
calculation yields
\begin{eqnarray}
q_1&=&q_0+hp_0-{\lambda\over2}h^2 q_0^3+\dots,\nonumber\\
p_1&=&p_0-\lambda h q^3_0-{3\over2}\lambda
h^2q_0p_0q_0+\dots.\label{exp}
\end{eqnarray}
We can define Fock-space creation and annihilation operators in
terms of the initial-time operators
\begin{equation}
q_0=\gamma{(a+a^\dagger)\over\sqrt{2}},\quad p_0={(a-a^\dagger)
\over i\sqrt{2}
\gamma},
\label{aadag}
\end{equation}
which satisfy
\begin{equation}
[a,a^\dagger]=1.
\end{equation}
Here we have introduced an arbitrary variational parameter
$\gamma$,
which represents the width of the corresponding harmonic oscillator
states.
The Fock-space states (harmonic oscillator states) are created and
destroyed
by these operators:
\begin{equation}
|n\rangle={(a^\dagger)^n\over\sqrt{n!}}|0\rangle,
\label{hostates}
\end{equation}
which states are not energy eigenstates of the anharmonic
oscillator.
We can now take matrix elements in these states of the dynamical
operators
at lattice site 1, using (\ref{exp}):
\begin{eqnarray}
\langle 1|p_1|0\rangle&\approx&\langle 1|p_0|0\rangle
(1+i{3\over2} h\lambda
\gamma^4-{3\over4}h^2\lambda\gamma^2+\dots)\nonumber\\
&\approx&\langle 1|p_0|0\rangle(1+i\omega h-{1\over2}\omega^2
h^2+\dots),
\end{eqnarray}
and
\begin{eqnarray}
\langle 1|q_1|0\rangle&\approx&\langle 1|q_0|0\rangle
(1+i{h\over\gamma^2}  -{3\over4}h^2\lambda\gamma^2+\dots)
\nonumber\\
&\approx&\langle 1|q_0|0\rangle(1+i\omega h-{1\over2}\omega^2
h^2+\dots),
\label{q1}
\end{eqnarray}
where we have assumed approximately exponential dependence
on the energy difference $\omega$.  Equating the coefficients
of the terms through order $h^2$ constitutes four
equations in two unknowns. These equations are consistent and yield
\begin{equation}
\omega={3\over2}\lambda\gamma^4={1\over\gamma^2},
\end{equation}
so the energy difference between the ground state and the first
excited
state is approximately
\begin{equation}
\omega=\left({3\over2}\lambda\right)^{1/3}\approx1.145\lambda^{1/3}
\end{equation}
which is only 5\% higher than the exact result
$E_{01}=1.08845\lambda^{1/3}$.
A similar calculation using quadratic finite elements ($r=2$) reduces
the
error to 0.5\%.

Numerical results for a large number of energy differences can also
be obtained by taking the discrete Fourier transform of the time
sequence $\{\langle0|q_n|1\rangle\}$.  For example, for 1000 finite
elements, energy differences are computed at the 2--3\% level
\cite{bengreen}.

\section{The Time-Evolution Operator}
Because the canonical commutation relations are preserved at each
lattice
site, we know that there is a unitary time evolution operator that
carries dynamical variables forward in time:
\begin{equation}
q_{n+1}=Uq_nU^{\dagger},\quad p_{n+1}=Up_nU^\dagger,
\end{equation}
 For the system described by the continuum Hamiltonian (\ref{ham}) in
the
linear finite-element scheme, we have found \cite{dtqm}
the following formula for
$U$:
\begin{equation}
U=e^{ihp_n^2/4}e^{ihA(q_n)}e^{ihp_n^2/4},
\end{equation}
where\footnote{A misprint occurs in (2.21b) of \cite{review}.}
\begin{equation}
A(x)={2\over h^2}[x-g^{-1}(4x/h^2)]^2+V(g^{-1}(4x/h^2)),
\end{equation}
\begin{equation}g(x)={4\over h^2}x+V'(x).
\label{gee}
\end{equation}
The implicit nature of the finite-element prescription is evident
in the appearance of the inverse of the function $g$.

Given the time evolution operator, a lattice Hamiltonian may be
defined
by $U=\exp(ih{\cal H})$.  For linear finite elements ${\cal H}$
differs
from the continuum Hamiltonian by terms of order $h^2$.
For example,
\begin{mathletters}
\label{quadfe}
\begin{eqnarray}
V={1\over2}m^2q^2: \quad & & {\cal H}={2\over
mh}\tan^{-1}\left(mh\over2\right)
\left[{1\over2}p^2+{1\over2}m^2q^2\right],\label{hc}\\
V={\lambda\over3}q^3:\quad & & {\cal H}={1\over2}p^2+{1\over3}
\lambda q^3
+h^2\left[{\lambda\over12}pqp+p^3\right]+\dots,\\
V={\lambda\over4}q^4:\quad & & {\cal H}={1\over2}p^2+{1\over4}
\lambda q^4
+h^2\left[-{\lambda^2\over24}q^6-{\lambda\over8}qp^2q\right]+\dots.
\label{ahc}
\end{eqnarray}
\end{mathletters}
If one uses quadratic finite elements ${\cal H}$ differs from the
continuum Hamiltonian by terms of order $h^4$, etc.

\section{Matrix Elements of Dynamical Variables}
\label{matq}
Remarkably, it is not necessary to solve the equations of motion
to compute matrix elements of the dynamical variable.  Introduce
creation and annihilation operators as in (\ref{aadag}).  Then, in
terms
of harmonic oscillator states (\ref{hostates}) the following formula
is
easily derived \cite{bss} for a general matrix element of $q_1$:
\begin{eqnarray}
\langle
m|q_1|n\rangle&&=-{\gamma\over\sqrt{2}}(\sqrt{m}\delta_{n,m-1}
+\sqrt{n}\delta_{m,n-1})\nonumber\\
\mbox{}+{e^{-i\theta(m-n)}\over R\sqrt{\pi
2^{n+m}n!m!}}&&\int_{-\infty}^\infty
dz\,ze^{-g^2(z)/4R^2}g'(z)H_n(g(z)/2R)H_m(g(z)/2R),
\end{eqnarray}
where $g$ is given by (\ref{gee}), $H_n(x)$ is the $n$th Hermite
polynomial,
 and we have introduced the abbreviations
\begin{equation}
R^2={4\gamma^2\over h^4}+{1\over h^2\gamma^2},\quad e^{-i\theta}
={2\gamma\over Rh^2}+{i\over R h\gamma}.
\end{equation}
For the example of the harmonic oscillator, this formula gives for
the
ground state--first excited state energy difference $\omega=(2/ h)
\tan^{-1}(h/2)$, consistent with (\ref{hc}), while for the
 anharmonic oscillator (\ref{qaho})
if we expand in $h$ we obtain precisely the expansion
(\ref{q1}).

\section{Matrix Elements of the Time Evolution Operator}
\label{matu}
A similar formula can be derived for the harmonic oscillator matrix
elements of the time evolution operator.  (There is an error in the
formula printed in \cite{bss}.)
\begin{eqnarray}
\langle m|U|n\rangle&=&{1\over2R}{1\over\sqrt{\pi
2^{n+m}n!m!}}e^{-i(n+m+1)\theta}\nonumber\\
&\times&\int_{-\infty}^\infty
dz\,g'(z)H_n(g(z)/2R)H_m(g(z)/2R)e^{[ihV(z)+ih^3V'(z)^2/8
- -h^2g(z)^2e^{-i\theta}/8\gamma R]},
\label{umat}
\end{eqnarray}
which again is expressed in terms of $g$ not $g^{-1}$.

For the harmonic oscillator, where $V=q^2/2$, (\ref{umat}) gives for
the
ground-state energy
\begin{equation}
\langle 0|U|0\rangle=e^{i\omega_0h},\quad \omega_0=
{1\over h}\tan^{-1}{h\over2},
\end{equation}
which follows from (\ref{hc}).  For the general
 anharmonic oscillator,
$V=\lambda q^{2k}/2k$,
again
 we expand in powers of $h$, with the result, for the harmonic
oscillator ground state,
\begin{eqnarray}
\langle 0|U|0\rangle&=&
1+ih\lambda^{1/(k+1)}f(\alpha)-{h^2\over2}\lambda^{2/(k+1)}s(\alpha)
\nonumber\\
&\approx&1+i\omega_0h-{1\over2}\omega^2_0h^2+\dots,
\label{woo}
\end{eqnarray}
where $\alpha=\lambda\gamma^{2k+2}$ and
\begin{mathletters}
\begin{eqnarray}
f(\alpha)&=&{1\over4\alpha^{1/(k+1)}}\left(1+
{2\alpha\over k}{\Gamma(k+1/2)\over\Gamma(1/2)}\right)\\
s(\alpha)&=&{1\over16\alpha^{2/(k+1)}}\left(3-
4\alpha{2k-1\over k}{\Gamma(k+1/2)\over\Gamma(1/2)}+{4\alpha^2\over
k^2}{\Gamma(2k+1/2)\over\Gamma(1/2)}\right).
\end{eqnarray}
\end{mathletters}
This result  also derivable from (\ref{quadfe}), using
(\ref{aadag}),
but with considerably more labor.
Equating powers of $h$ in (\ref{woo}) gives us two equations, which
 are to be solved first for the dimensionless number
$\alpha$.
Once the number $\alpha$ is determined, the value of $\omega_0$ is
expressed as
\begin{equation}
\omega_0=\lambda^{1/(k+1)}f(\alpha).
\label{first}
\end{equation}

For a first estimate, we use only the $O(h)$ equation (\ref{first})
 and employ the ``principle of minimum sensitivity''
(PMS) \cite{pms}  that is, use the stationary value of
$\alpha$,
\begin{equation}
f'(\alpha)=0\Rightarrow \alpha={2^{k-1}\over(2k-1)!! }
\Rightarrow f(\alpha)={k+1\over4k}{[(2k-1)!!]^{1/(k+1)}\over
2^{(k-1)/(k+1)}}.
\end{equation}
Specific examples are
\begin{equation}
f(\alpha)=\left\{\begin{array}{cc}
0.4293,& k=2,\\
0.4639,&k=3,\\
0.5230,&k=4,\\
\end{array}\right.
\end{equation}
which are higher than the exact values \cite{bms} of
0.420805, 0.43493, and 0.46450 by about 2\%, 7\%, and 13\%,
respectively.
  In fact,
when we solve (\ref{woo}) for $\alpha$, that is solve
$f(\alpha)^2=s(\alpha)$,
 we find  complex values, for example, for $k=2$
\begin{equation}
\alpha={1\over2}\pm{i\over2\sqrt{3}}\Rightarrow
f(\alpha)=0.4178\mp0.0077i.
\label{wo1}
\end{equation}
The imaginary part is small, and the real part is only 0.7\% low.
The corresponding result for $k=3$ is
$f(\alpha)=0.4453\mp0.0352i$, where the real part is now only
2\% high.  However, for $k=4$, $f(\alpha)=0.5171\mp0.0713i$,
and the real part is still 11\% high.
The failure of (\ref{wo1}) to be real does not indicate any
breakdown of unitarity, but only that the one state
approximation is not exact.

\begin{figure}
  \centerline{\psfig{figure=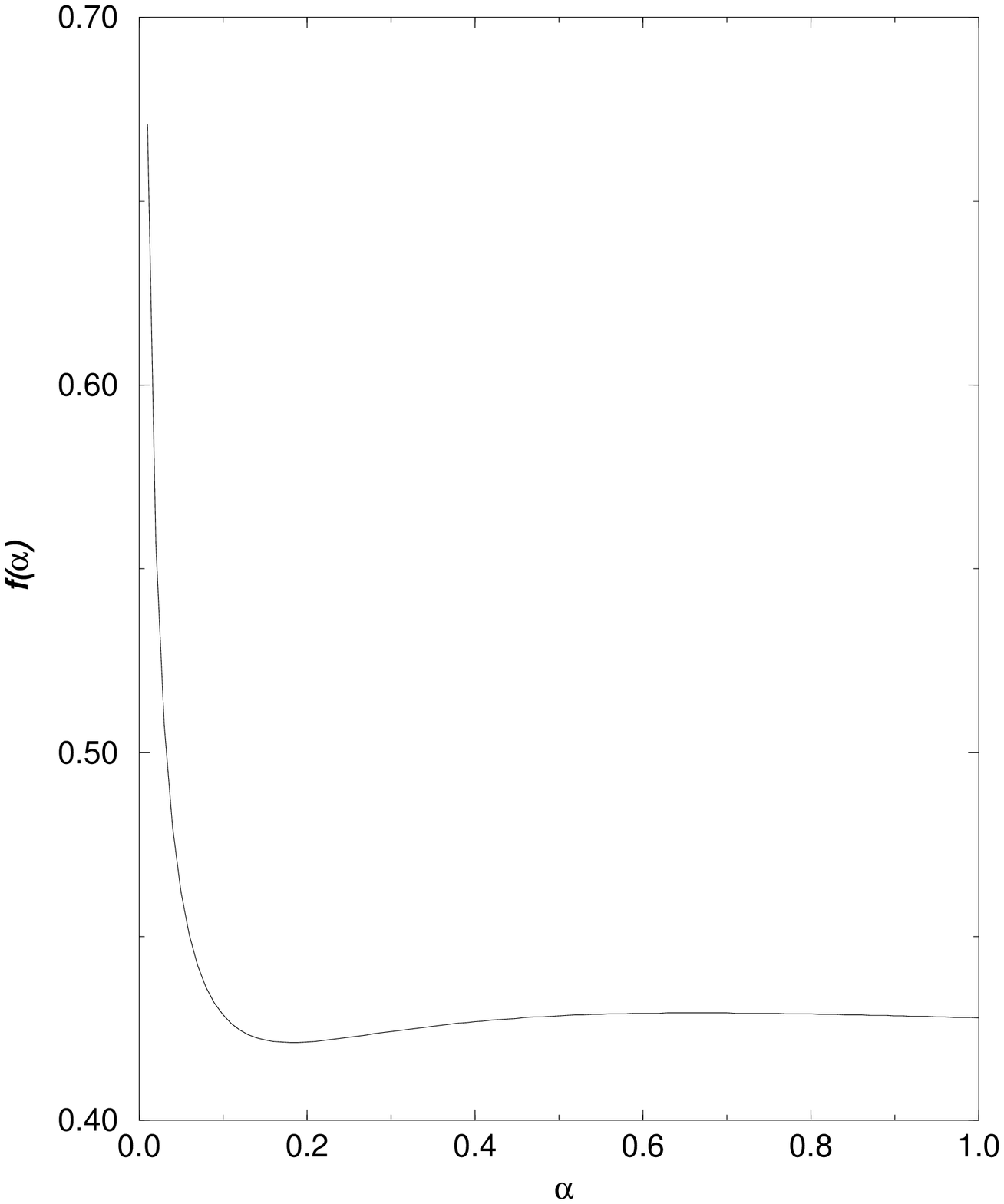,height=6.5in,width=5in}}
 \caption{Ground-state energy for the quartic anharmonic
oscillator as a function of
$\alpha=\lambda\gamma^6$, in the second approximation.
Here $\omega_0=\lambda^{1/3}f(\alpha)$, $f(\alpha)$ given by
(\protect\ref{omgood}).}
\label{fig1}
\end{figure}

We do much better by making a two-state approximation, where we must
diagonalize the $2\times2$ matrix
\begin{equation}
\left(\begin{array}{cc}
U_{00}&U_{02}\\
U_{20}&U_{22}\\
\end{array}\right).
\label{u}
\end{equation}
For $k=2$ we then find the following relation between
$\omega_{0,2}$ and
$\alpha=\lambda\gamma^6$:
\begin{equation}
\omega_{0,2}={\lambda^{1/3}\over16}\alpha^{-1/3}[12+21\alpha
\mp2\sqrt{3}(8+16\alpha+33\alpha^2)^{1/2}],
\label{omgood}
\end{equation}
which, for the $-$ sign, is plotted in Fig.~\ref{fig1}.
This  graph shows that the ground-state energy is very insensitive
to the value of $\alpha$ for a broad range of values.
  The principle of minimum sensitivity
gives
\begin{equation}
\omega_0=0.42124\lambda^{1/3},
\end{equation}
in  spectacular agreement with the exact result,
being only 0.1\% high,
while it gives a value for the third state,
$\omega_2=2.992\lambda^{1/3},$ accurate to 1\%.
(The exact value is $2.959\lambda^{1/3}$ \cite{bo}.)
Solving for $\alpha$ from the eigenvalues of (\ref{u}) gives even
better results:
\begin{equation}
\omega_0=\lambda^{1/3}(0.42054\pm2\times 10^{-6}i), \quad
\omega_2=\lambda^{1/3}(2.9433\pm0.0220i),
\end{equation}
where the ground state energy is now low by 0.06\%, the imaginary
part
being negligible.  The real part of the energy of the third state
is low by only 0.5\%.  These results for the ground state are much
better than those resulting from the WKB approximation \cite{bo}.

The corresponding PMS values for the $k=3$ and $k=4$  oscillators
are similarly impressive: $\omega_0=0.43913\lambda^{1/4}$ (+0.9\%)
and $\omega_0=0.47718\lambda^{1/5}$ (+2.7\%), respectively.
Using the $O(h^2)$ data to compute $\alpha$ gives truly outstanding
agreement:
\begin{mathletters}
\begin{eqnarray}
k=3:\quad \omega_0&=&(0.43284\pm0.00259i)\lambda^{1/4}\\
\omega_2&=&(3.4532\pm0.1271i)\lambda^{1/4}\\
k=4:\quad \omega_0&=&(0.46347\pm0.01463i)\lambda^{1/5}\\
\omega_2&=&(4.0186\pm0.3081i)\lambda^{1/5}
\end{eqnarray}
\end{mathletters}
where the real parts of the ground-state energies are now low by
0.5\% and 0.2\%, respectively.

\section{Wavefunctions}
\label{wf}
\begin{figure}
\centerline{\psfig{figure=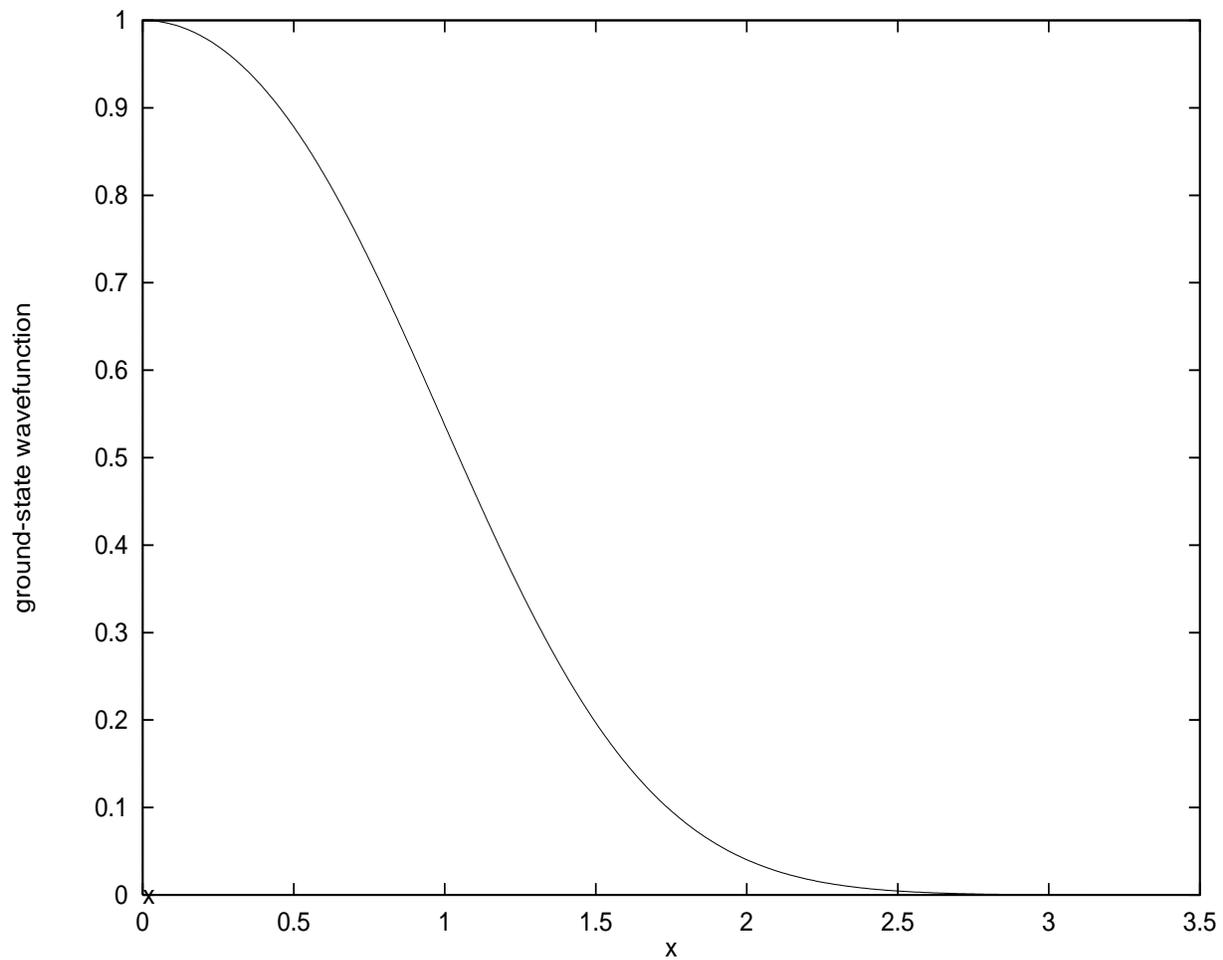,height=5in,width=6.5in,angle=270}}
\caption{Ground-state wavefunction in the
two-state approximation.}
\label{fig2}
\end{figure}

\begin{figure}
\centerline{\psfig{figure=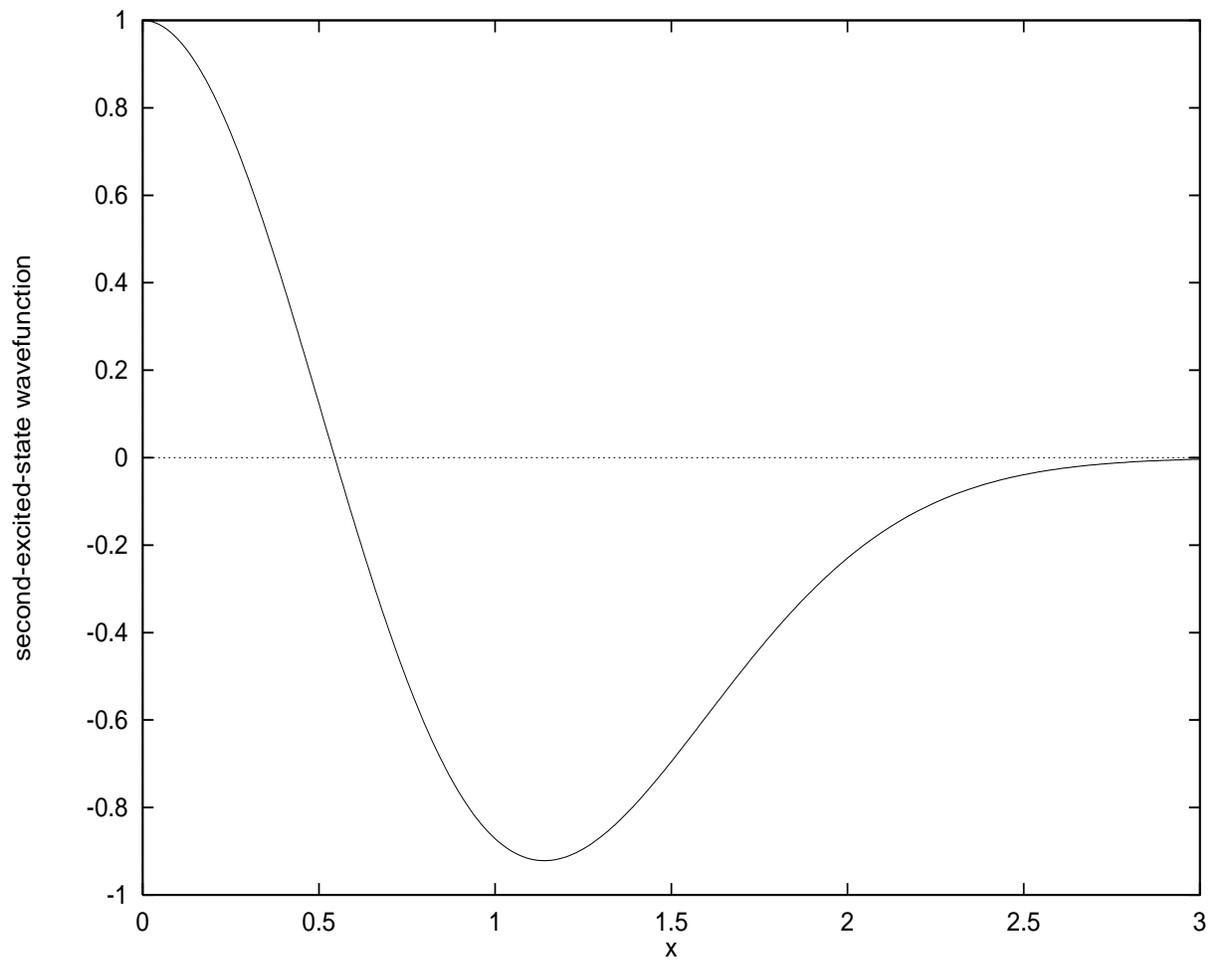,height=5in,width=6.5in,angle=270}}
\caption{Second-excited state  wavefunction in the
two-state approximation.}
\label{fig3}
\end{figure}

In the process of diagonalizing (\ref{u}) we find the corresponding
wavefunctions in the two-state approximation,
that is, the wavefunctions are taken to be linear combinations
of $n=0$ and $n=2$ harmonic oscillator states of width $\gamma$.
  The real parts of these wavefunctions are plotted in
Figs.\ \ref{fig2} and \ref{fig3}. (The imaginary parts amount
to only
2\% for the ground-state wavefunction and 5\% for the excited-state
wavefunction.)
These are normalized to unity at
the origin to facilitate comparison with \cite{bo}.

When these are compared with the exact wavefunctions given in
\cite{bo}, we see that the approximate
ground-state wavefunction is nearly
indistinguishable from the exact one, and is much better
that the WKB wavefunction given there.  The excited-state
wavefunction is quite good, but deviates slightly from the
exact wavefunction, and in particular, the minimum at $x\approx1.1$
should be about 10\% deeper.  This deviation is not surprising,
since the exact excited-state
 wavefunction must contain a substantial mixing with the $n=4$
harmonic oscillator state. The error in the excited-state
wavefunction
is also manifested  in that fact that in this approximation the
two wavefunctions are not quite orthogonal, the magnitude of the
overlap being 5\%.

\section{Conclusions}

The simple calculations given here for quantum-mechanical
anharmonic oscillators are the beginning of a program to develop
use of lattice Hamiltonian techniques to explore gauge theories
in the finite-element context. The  numerical results presented
in Sec.~\ref{matu} and \ref{wf}
also hold in the continuum, by virtue of (\ref{quadfe}),
but the calculations are much more tedious without the
use of the finite-element formula (\ref{umat}).  It is in
two or more space-time dimensions that the essential nature of
the lattice in such calculations comes into play
\cite{mmsb,mmsb2,bm}.
 The high accuracy contrasted with
the simplicity of the approach leads us to expect that we can
extract spectral information, anomalies, and symmetry breaking
from an examination of the time-evolution operator in gauge
theories.

\section*{Acknowledgments}
KAM  thanks the US Department of Energy, the UK PPARC, and the
University
of Oklahoma
College of Arts and Sciences
 for partial financial support of this
research.

\end{document}